\documentclass[sigconf,natbib=true, nonacm]{acmart}
\AtBeginDocument{%
  }

\usepackage{float}
\usepackage{booktabs,subcaption}

\begin{document}

\title{Efficient Long-Document Reranking via Block-Level Embeddings and Top-k Interaction Refinement}


\author{Minghan Li}
\orcid{0000-0002-1041-8887}
\email{mhli@suda.edu.cn}
\affiliation{%
  \institution{Soochow University}
  \city{Suzhou}
  \postcode{215006}
  \country{China}
}

\author{Eric Gaussier}
\orcid{0000-0002-8858-3233}
\email{Eric.Gaussier@imag.fr}
\affiliation{%
  \institution{
  Univ Grenoble Alpes, CNRS, Grenoble INP, LIG
}
  \city{Grenoble}
  \postcode{38401}
  \country{France}
}

\author{Guodong Zhou}
\orcid{0000-0002-7887-5099}
\email{gdzhou@suda.edu.cn}
\affiliation{%
  \institution{Soochow University}
  \city{Suzhou}
  \postcode{215006}
  \country{China}
}

\renewcommand{\shortauthors}{Li et al.}

\begin{abstract}
Dense encoders and LLM-based rerankers struggle with long documents: single-vector representations dilute fine-grained relevance, while cross-encoders are often too expensive for practical reranking. We present an efficient long-document reranking framework based on block-level embeddings. Each document is segmented into short blocks and encoded into block embeddings that can be precomputed offline. Given a query, we encode it once and score each candidate document by aggregating top-$k$ query--block similarities with a simple weighted sum, yielding a strong and interpretable block-level relevance signal. To capture dependencies among the selected blocks and suppress redundancy, we introduce Top-$k$ Interaction Refinement (TIR), a lightweight setwise module that applies query-conditioned attention over the top-$k$ blocks and produces a bounded residual correction to block scores. TIR introduces only a small number of parameters and operates on top-$k$ blocks, keeping query-time overhead low. Experiments on long-document reranking benchmarks (TREC DL and MLDR-zh) show that block representations substantially improve over single-vector encoders, and TIR provides consistent additional gains over strong long-document reranking baselines while maintaining practical reranking latency (e.g., on TREC DL'23, NDCG@10 improves from 0.395 to 0.451 at the same block budget $k{=}65$ ($\leq$4095 tokens)). The resulting model supports interpretability by exposing which blocks drive each document's score and how refinement redistributes their contributions.
\end{abstract}



\maketitle

\section{Introduction}
Large language models (LLMs) have demonstrated strong capabilities across a broad range of NLP tasks \cite{zhao2023survey,dubey2024llama}. 
In web search and information retrieval (IR), they are increasingly used as representation-based encoders \cite{zhao2024dense} to compute embeddings for queries and documents, enabling similarity-based ranking and supporting retrieval-augmented generation (RAG) pipelines \cite{wang2023improving,ma2024fine,gao2023retrieval}. 
When applied to long and heterogeneous web documents, the common practice is to generate a single coarse-grained embedding after truncating the input. 
This practice often mixes relevant and irrelevant content, which dilutes fine-grained signals and reduces both retrieval effectiveness and interpretability.

\begin{figure}[t] 
  \hfill 
    \centering
    \includegraphics[width=\linewidth]{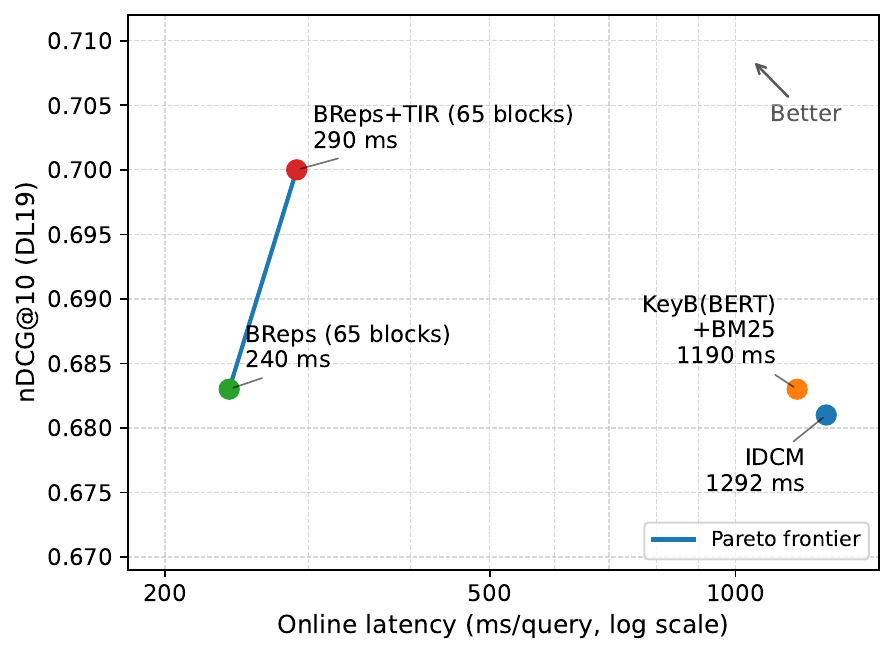}
\caption{Effectiveness--efficiency trade-off on TREC DL'19 (Doc). We plot nDCG@10 against \emph{online} reranking latency (ms/query) for reranking the top-100 candidates (document-side block embeddings are precomputed offline), measured on a single NVIDIA A100 GPU (batch size 8).
}
    \label{fig:ndcg-length}
\end{figure}

This paper revisits document representation granularity for long-document reranking.
We argue that a single, coarse document embedding is not the ideal unit for LLM-based ranking, and advocate a fine-grained, block-level alternative.
We propose \textbf{BReps}, a representation-based \emph{long-document reranking} framework using a decoder-only LLM as the encoder: it segments each document into informative blocks, encodes blocks independently, and scores each candidate by aggregating the top-$k$ query--block similarities with a weighted sum.
To further improve calibration among the top-$k$ blocks, we introduce \textbf{BReps-TIR}, a query-conditioned interaction module that performs setwise attention over the selected blocks and refines block scores before aggregation.
Because document-side block embeddings are computed offline, online reranking remains efficient while preserving localized evidence.

Empirically, our approach attains strong effectiveness at low cost.
As highlighted in Figure~\ref{fig:ndcg-length}, with only 20k training triplets and a LoRA-tuned Gemma2-2B encoder, BReps and BReps-TIR achieve a favorable effectiveness--efficiency trade-off on long-document reranking.
In particular, they outperform the single-vector LLM reranker RepLLaMA, and are substantially faster than interaction-heavy long-document rerankers such as IDCM and KeyB under the same fixed-candidate protocol.
These results provide concrete evidence that fine-grained block-level representations with small LLMs can be both accurate and efficient.

Beyond accuracy, BReps offers practical advantages. First, it requires no changes to the vector-store interface: block embeddings can be indexed as standard vectors, while top-$k$ interactions are computed at query time. Second, the block-level design naturally supports interpretability by highlighting which segments dominate the final score, which is valuable in domains such as legal case retrieval and scholarly search. 
Third, the offline computation of block-level document embeddings reduces online latency and stabilizes serving costs compared to single-shot long-input encoders.

Our contributions are summarized as follows:
\begin{enumerate}
  \item We propose \textbf{BReps}, a fine-grained, representation-based \emph{long-document reranking} framework that encodes each document into block-level embeddings and aggregates the top-$k$ query--block similarities for robust document-level ranking.
  \item We introduce \textbf{BReps-TIR}, a lightweight query-conditioned interaction module that performs setwise attention over the top-$k$ selected blocks to refine their scores before aggregation, improving calibration with modest online overhead.
  \item We conduct extensive experiments on long-document \emph{reranking} benchmarks (including TREC DL) showing that, with only 20k training triplets and a LoRA-tuned Gemma2-2B encoder, our models consistently improve effectiveness over RepLLaMA and remain substantially more efficient than cross-encoder-based long-document rerankers such as IDCM and KeyB under the same fixed-candidate protocol (Fig.~\ref{fig:ndcg-length}).
\item We provide thorough ablations on segmentation, aggregation, and key hyperparameters, and show that \textbf{TIR} is a lightweight yet effective refinement: it yields consistent gains over BReps while adding only modest online overhead. We further present an efficiency analysis that decomposes offline representation computation and online reranking latency.
\end{enumerate}

The remainder of this paper is organized as follows. Section~\ref{sec:related} reviews related work. Section~\ref{sec:method} details BReps and the TIR refinement. Section~\ref{sec:experi} presents experimental setups, main results, and ablations. Section~\ref{sec:conclusion} concludes the paper.

\section{Related Work}
\label{sec:related}

\subsection{Representation-based vs.\ Interaction-based Models}
The introduction of Transformer and BERT \cite{vaswani2017attention,devlin-etal-2019-bert} triggered large improvements across IR tasks. Neural retrieval methods are usually grouped into three categories: interaction-based, representation-based, and late-interaction.  
Interaction-based models, such as BERT re-ranking \cite{nogueira2019passage}, concatenate query and document tokens and feed them into a cross-encoder. They capture rich matching signals and consistently achieve strong accuracy, but their online cost is prohibitive since all query–document pairs must be processed jointly.  
Representation-based models, also known as dense retrieval \cite{huang2013learning,karpukhin2020dense,reimers-gurevych-2019-sentence}, encode queries and documents independently into single vectors in a shared space, and compute similarity with dot product or cosine similarity. They enable pre-computation and scalable search via ANN, but their coarse single-vector representations often struggle to capture subtle relevance signals. Many follow-ups focus on mining hard negatives or addressing domain adaptation issues \cite{xiong2020approximate,wang2022gpl,li2024domain,zhan2021optimizing}.  
Late-interaction models strike a middle ground. ColBERT \cite{khattab2020colbert} encodes queries and passages separately but retains token-level embeddings, enabling fine-grained max-sim interactions at query time. ICLI \cite{li2022bert} learns passage-level embeddings offline and speeds up block selection at retrieval. These approaches balance effectiveness and efficiency, but they generally rely on bidirectional encoders and are not directly compatible with current decoder-only LLMs. Our work shares with late-interaction the idea of multi-representations, but differs by generating block-level LLM embeddings for long documents, which can be precomputed offline and then refined efficiently online.

\begin{figure*}[t]
\centering
\includegraphics[width=0.95\textwidth]{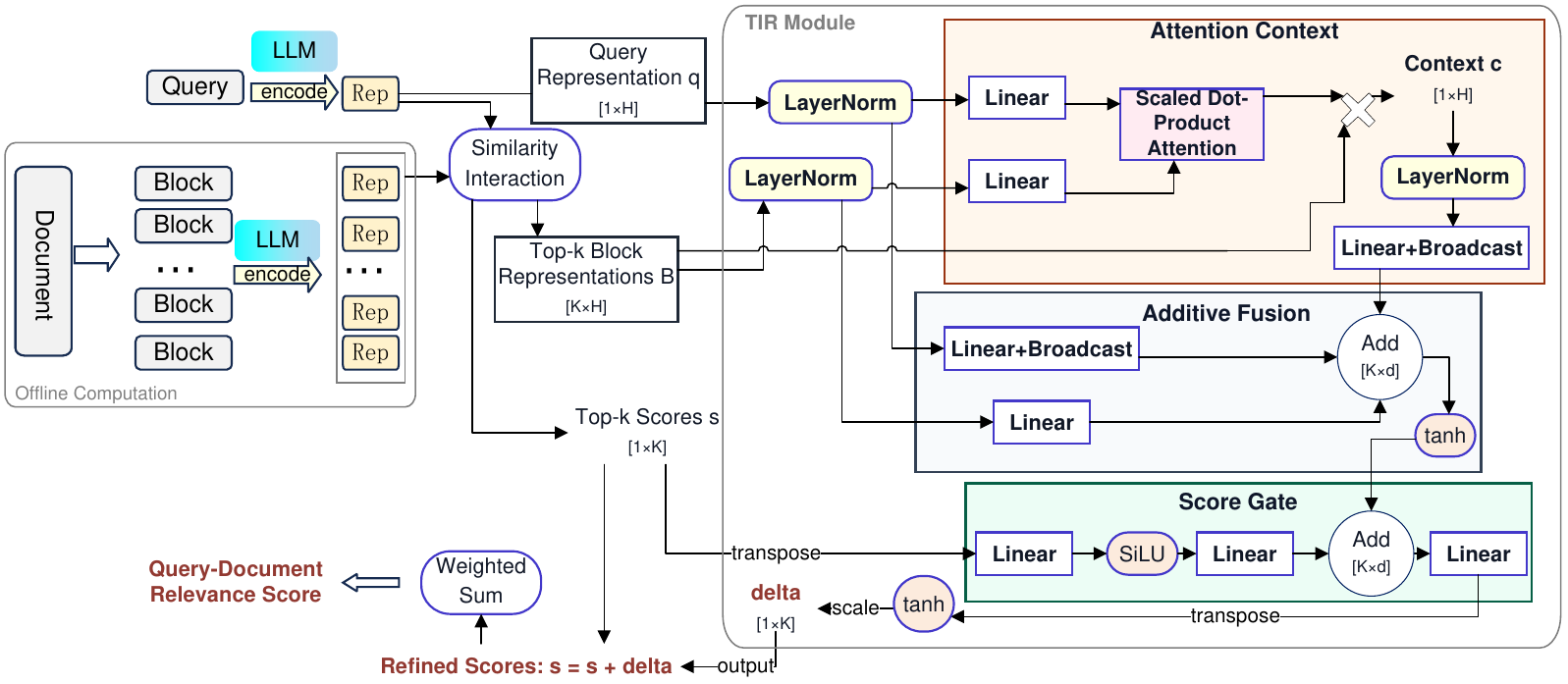} 
\caption{\textbf{Overview of BReps-TIR.} \textbf{Offline:} segment each document into blocks and precompute block embeddings. \textbf{Online:} encode the query once, compute query--block similarities for each candidate, select top-$k$ blocks, and aggregate their scores with a descending-weighted sum to obtain a document score. \textbf{TIR:} a lightweight query-conditioned setwise module that attends over the top-$k$ blocks and applies a bounded residual correction to calibrate block scores before aggregation.}
\label{fig:arch}
\end{figure*}

\subsection{Neural Long-Document Retrieval}
Long documents pose additional challenges: lengths far exceeding typical PLM input windows, and relevance signals scattered across sections. Existing neural methods fall into three broad directions.  
First, sparse-attention mechanisms such as QDS-Transformer \cite{jiang2020long} and Longformer \cite{beltagy2020longformer} prune self-attention patterns to reduce quadratic complexity. While they improve efficiency, they require custom kernels or special frameworks \cite{chen2018tvm,gray2017gpu,li2023power}, making deployment costly.  
Second, passage aggregation approaches split documents into shorter segments and then aggregate passage-level signals. MaxP or FirstP \cite{dai2019deeper} use simple pooling, while PARADE \cite{li2023parade} employs transformer layers to combine passage representations. These methods can capture distributed relevance but remain memory-hungry and slow since each query–passage pair must be scored online.  
Third, key passage or block selection approaches explicitly identify the most relevant segments before computing document scores. KeyBLD and its extensions \cite{li2021keybld,li2023power} as well as IDCM \cite{DBLP:conf/sigir/HofstatterMZCH21} achieve strong accuracy while reducing cost. ICLI \cite{li2022bert} further enables fast retrieval by pre-storing passage vectors. These strategies show better efficiency but still depend on interaction-based scoring or additional learning pipelines. Other recent work explored fine-grained knowledge distillation \cite{zhou2024fine} or multi-view indexing \cite{dong2024mc}, yet still compress documents into coarse single vectors or require extra features such as keywords and summaries.

Recently, LLMs with extended context windows have been explored for IR. RankingGPT \cite{zhang2024two} and RankLLaMA \cite{ma2024fine} use cross-encoder style relevance estimation with LLMs, achieving strong accuracy but incurring very high inference cost. RepLLaMA \cite{ma2024fine} instead encodes the whole document into a single LLM vector, which is efficient but coarse. Analyses show that LLMs are prone to the ``lost in the middle'' effect \cite{liu2024lost}, where information in the middle of long inputs is underutilized. Moreover, complexity of transformer-based LLMs grows roughly quadratically with input length, even with efficiency improvements like grouped-query attention \cite{ainslie2023gqa}.  

Beyond these, a new line of embedding-focused baselines has emerged.
LongEmbed (E5-Base-4K) \cite{zhu2024longembed} extends embedding models to 4k context length, enabling better handling of long inputs while retaining efficiency. BGE-M3 \cite{chen2024m3} introduces a multi-lingual, multi-functionality, and multi-granularity framework, aiming to unify dense retrieval, classification, and clustering with one versatile embedding model. These models highlight the potential of embedding-driven retrieval for both efficiency and generalization.  

Our method, BReps, differs in two key aspects. First, instead of compressing each long document into a single coarse vector, we retain multiple fine-grained block-level embeddings produced by an LLM, which better preserves local relevance signals. Second, BReps requires substantially fewer training data—only about 20k labeled examples—to surpass not only smaller encoder-based models but in some cases even 7B-level embedding models trained with millions of synthetic examples. By combining block-level offline representations with a lightweight Top-$k$ Interaction Refinement module, BReps achieves a more favorable trade-off between accuracy and efficiency in long-document retrieval.

\section{Method}
\label{sec:method}

As discussed above, our proposed approach leverages fine-grained block representations instead of a single, global representation for the whole document. The architecture involves three main components: (1) document segmentation and block encoding, where each document is divided into smaller units and encoded independently; (2) initial scoring and top-K block selection, which provides a fine-grained relevance estimation; and (3) a lightweight Top-K Interaction Refinement (TIR) module, where the query interacts with the top-ranked blocks to produce refined scores by applying residual adjustments.
The overall architecture is illustrated in Fig.~\ref{fig:arch}, and each component is described in detail below.

\subsection{Block Representations for Long Document}
This enables fine-grained way to represent a long document, with dozens or hundreds of representations. 
The first step is to segment each document into a sequence of blocks (the top-left part of Fig.~\ref{fig:arch}), which are relatively short and contain relatively useful information.
In this paper, we adopt the CogLTX block segmentation approach \cite{ding2020cogltx}. It uses dynamic programming, and aims to segment a document into blocks according to punctuation marks which have different weights. The maximum tokens for a block is 63. This approach prefers a separation with a period, an exclamation mark or a question mark. 
This segmentation approach suits the intuition that a block should be relatively short and homogeneous. 
After segmenting a long document into blocks, we generate representations for them using the same approach as RepLLaMA \cite{ma2024fine}, which leverages decoder-only LLMs to represent passages. 
We use a maximum number of $n$ blocks to efficiently accommodate the GPU's memory constraints during training and retain only the fits $n$ blocks of each document, which are represented as:
\begin{equation*}
  f\_{block_j} = \text{passage: }t_{1}^{j}\text{\textvisiblespace}t_{2}^{j}\text{\textvisiblespace}...\text{\textvisiblespace}t_{i}^{j}\texttt{</s>}, \, 1 \le j \le n
  \end{equation*}
where $t_{1}^{j}\text{\textvisiblespace}t_{2}^{j}\text{\textvisiblespace}...\text{\textvisiblespace}t_{i}^{j}$ are the tokens of the $j^{th}$ block.
They are then fed to an LLM simultaneously to obtain the block representations:
\begin{align}
\label{eq:block-repres}
  V_{b_j} = \text{LLM}(f\_{block_j})[-1],  \, 1 \le j \le n. 
\end{align}

\subsection{Fine-grained Interaction between Queries and Block Representations}
\label{sec-method-interaction}

As queries are usually short, we do not segment them and fed them directly to the query encoder, which is different with the block encoder. However, for simplicity, we used in this work the same encoder for queries and blocks, which is a standard paradigm of bi-encoder models.
The query input format and encoding parallel the ones of the blocks:
\begin{equation*}
  f\_{query} = \text{query: }t_{1}\text{\textvisiblespace}t_{2}\text{\textvisiblespace}...\text{\textvisiblespace}t_{q}\texttt{</s>},
  \end{equation*}
where $t_{1}\text{\textvisiblespace}t_{2}\text{\textvisiblespace}...\text{\textvisiblespace}t_{q}$ are the tokens of the query. 

Furthermore:
\begin{align}
\label{eq:query-repres}
  V_q = \text{LLM}(f\_{query})[-1]. 
\end{align}

Firstly, each block representation is matched to the query representation using the standard cosine similarity in a way similar to RepLLaMA \cite{ma2024fine}, which yields a relevance score for each block (shown as "Similarity Interaction" in Fig.~\ref{fig:arch}):
\begin{align}
\label{eq:block-score}
  S_{q,{b_j}} = \cos({V_q,V_{b_j}})/0.01, \, 1 \le j \le n. 
\end{align}

After calculating each query-block relevance score, we need to identify the overall query-document score. 
To simulate human judgment and following previous work, as ICLI \cite{li2022bert}, we aggregate the relevance scores of the top $k$ relevant blocks to obtain the document relevance score. The aggregation retained here is the one proposed in \cite{li2022bert}, which amounts to a weighted sum of the relevance scores of the top $k$ blocks, the weights being higher for more relevant blocks than for less relevant ones. Denoting by $S_{q,b}^{(i)}$ the score of the $i^{th}$ most relevant block, the initial document score of a document $d$ is given by:
\begin{align}
\label{eq:doc-score}
S_{q,d} = [S_{q,b}^{(1)}, ..., S_{q,b}^{(k)}]\times W^T , 
\end{align}
where $W=[w_1, ..., w_k]$ is a weight vector such that $w_1 \ge w_2 \ge ... \ge w_k$. The vector $W$ can either be set or learned from data, in which case one may drop the ranking constraints on the weights. This is the proposed default version BReps model.

\subsection{Top-$k$ Interaction Refinement (\textsc{TIR})}
\label{sec:tir}

\textbf{Motivation.}
Eq.~\ref{eq:doc-score} aggregates the top-$k$ block scores independently, which may
(i) \emph{over-count redundancy} when multiple blocks repeat the same evidence, and
(ii) \emph{underestimate complementarity} when several medium-quality blocks jointly cover different query facets.
To address this, we introduce \textsc{TIR}, a lightweight query-conditioned module (Fig.~\ref{fig:arch}, right) that models setwise interactions among the selected blocks and refines their scores via a \emph{bounded} residual.
Importantly, \textsc{TIR} operates only on the top-$k$ blocks and \emph{does not} re-encode documents.

\paragraph{Notation.}
Let the query representation be $V_q \in \mathbb{R}^{H}$ and the selected top-$k$ block representations be
$B=[b_1;\ldots;b_k]\in\mathbb{R}^{k\times H}$, sorted by their original block scores $s\in\mathbb{R}^{k}$.
We use a small projection dimension $d \ll H$ and apply LayerNorm row-wise:
$V_q^{(n)}=\mathrm{LN}(V_q)$ and $B^{(n)}=\mathrm{LN}(B)$.

\paragraph{Query-to-block attention and setwise context.}
We compute a single-head attention from the query to the $k$ blocks:
\begin{equation}
\label{eq:tir-attn}
\hat q = W_q^{\mathrm{att}} V_q^{(n)} \in \mathbb{R}^{d},\quad
\hat B = B^{(n)} (W_b^{\mathrm{att}})^\top \in \mathbb{R}^{k\times d},
\end{equation}
\begin{equation}
\alpha = \mathrm{softmax}\!\left(\tfrac{\hat B \hat q}{\sqrt{d}\,\tau_{\mathrm{att}}}\right)\in\mathbb{R}^{k},\quad
c = \mathrm{LN}\!\left(\sum_{i=1}^{k}\alpha_i\, b_i^{(n)}\right)\in\mathbb{R}^{H},
\end{equation}
where $W_q^{\mathrm{att}},W_b^{\mathrm{att}}\in\mathbb{R}^{d\times H}$ and $\tau_{\mathrm{att}}>0$.
The context $c$ summarizes query-conditioned evidence across the selected set, helping identify redundancy and complementarity.

\paragraph{Per-block fusion.}
We project the query, each block, and the setwise context into a shared $d$-dimensional space:
\[
q' = W_q V_q^{(n)},\quad b'_i = W_b b_i^{(n)},\quad c' = W_c c,
\qquad W_q,W_b,W_c\in\mathbb{R}^{d\times H}.
\]
Stack $b'_i$ as $B'=[b'_1;\ldots;b'_k]\in\mathbb{R}^{k\times d}$ and broadcast $q'$ and $c'$ across rows:
\[
Z = \tanh\!\big(\mathbf{1}_k (q')^\top + B' + \mathbf{1}_k (c')^\top\big) \in \mathbb{R}^{k\times d}.
\]

\paragraph{Score-gated residual calibration.}
Let $s\in\mathbb{R}^{k}$ denote the base scores of the selected top-$k$ blocks (sorted by $s$; we keep this top-$k$ set and order fixed).
To incorporate the original score ordering, we map $s$ through a small MLP to obtain a score gate $G\in\mathbb{R}^{k\times d}$ and add it to $Z$:
\[
G=\mathrm{MLP}(s)\in\mathbb{R}^{k\times d},\qquad
Z \leftarrow Z + G.
\]
A linear head then predicts a per-block residual, which is bounded by $\tanh$ and scaled by $\gamma$:
\begin{equation}
\label{eq:tir-residual}
\delta_{\mathrm{raw}} = (W_v Z^\top)^\top \in \mathbb{R}^{k},\quad
\delta = \gamma \tanh(\delta_{\mathrm{raw}})\in\mathbb{R}^{k}.
\end{equation}
We then obtain calibrated scores by $s' = s + \delta$, where $W_v \in \mathbb{R}^{1\times d}$.

\paragraph{Refined document score.}
TIR outputs calibrated block scores $s'$. We then apply the same top-$k$ aggregation as Eq.~\ref{eq:doc-score}:
\begin{equation}
\label{eq:tir-refined-score}
S_{q,d}=\sum_{i=1}^{k} w_i\, s'_{(i)}
= \underbrace{\sum_{i=1}^{k} w_i\, s_{(i)}}_{\text{aggregation}}
+ \underbrace{\sum_{i=1}^{k} w_i\, \delta_{(i)}}_{\text{refinement}}.
\end{equation}

\section{Experiments}
\label{sec:experi}

\subsection{Datasets}
\label{sec:datasets}

We evaluate on four long-document benchmarks: TREC 2019 DL (MS MARCO v1, document ranking)~\cite{craswell2020overview}, TREC 2022/2023 DL (MS MARCO v2)~\cite{craswell2025overview,craswell2025overviewtrec2023deep}, and MLDR-zh.\footnote{\url{https://huggingface.co/datasets/Shitao/MLDR}}
These corpora differ in domain, language, and document length, covering multiple length regimes (from $\sim$1.6k tokens on DL19 to $>$9k tokens on MLDR-zh).
We use the official test queries and evaluation splits.

Table~\ref{tab:datasets} summarizes the dataset statistics (Gemma-2-2B tokenizer).
We provide further preprocessing details in Section~\ref{sec:impl-details}. 

\begin{table}[htbp]
  \centering
  \caption{Dataset statistics and evaluation splits.}
  \label{tab:datasets}
  \resizebox{\columnwidth}{!}{%

\begin{tabular}{@{}lllll@{}}
\toprule
Dataset
&genre 
&\#documents &\#test query &Avg. Num. of Tokens \\ \midrule
TREC 2019 DL doc & Web Documents &3,213,835& 49 & 1670\\
TREC 2022 DL doc & Web Documents &11,959,635& 76 & 3542\\
TREC 2023 DL doc & Web Documents &11,959,635& 82 & 3542\\
MLDR-zh &Wikipedia + Wudao&200000& 800 &  9030\\
 \bottomrule
\end{tabular}

  }
\end{table}

\subsection{Baselines and Evaluation}
\label{sec:baselines-eval}

\paragraph{Task definition.}
Our goal is long-document \emph{reranking} under fixed candidate sets: official top-100 for TREC DL and the official candidates for MLDR-zh.
All compared methods rerank the \emph{same} candidates for each query.
For models with limited input length (e.g., 512 tokens), we apply truncation to fit the maximum window; block-based methods operate on segmented blocks and aggregate block-level evidence into a document score.

\paragraph{Baselines.}
We compare against four groups of baselines:

\noindent\textbf{(1) Full-document single-pass scoring (non-segmented).}
We include RepLLaMA~\cite{ma2024fine} as a strong dense reranker using a single document embedding, and BM25~\cite{robertson2009probabilistic} as a lexical baseline for reranking within the same candidate set. BM25 results are obtained with Anserini \cite{yang2018anserini} using default settings.

\noindent\textbf{(2) Short-context dense and late-interaction rerankers (512-token window).}
We train BERT-dot~\cite{reimers-gurevych-2019-sentence} and ColBERT~\cite{khattab2020colbert}.
Since these models are typically limited to a 512-token window, they rely on truncation when applied to long documents, which may reduce coverage in the reranking setting.

\noindent\textbf{(3) Long-context embedding models.}
We further compare with embedding models that support longer contexts:
E5-Base-4K (LongEmbed)~\cite{zhu2024longembed}
and BGE-M3~\cite{chen2024m3}.

\noindent\textbf{(4) Segmentation-and-aggregation long-document rerankers.}
To match our block-based paradigm, we include IDCM~\cite{DBLP:conf/sigir/HofstatterMZCH21} and KeyB(BERT)$_{\text{BM25}}$ \cite{li2023power}. We use the official IDCM checkpoint; since it does not provide a Chinese model, we do not report IDCM results on MLDR-zh.
For KeyB, we use the authors' code and recompute the BM25 IDF statistics on the corresponding training corpus for each language.

\paragraph{Evaluation protocol.}
We follow the official evaluation setup for each benchmark (Section~\ref{sec:datasets}).
For TREC DL (DL19/22/23), we report NDCG@10 as the primary metric and also include MAP.
For MLDR-zh, we rerank the official candidate set (one positive and seven negatives per query); accordingly, we report P@1, MAP, and NDCG@8.

\subsection{Implementation Details}
\label{sec:impl-details}

\noindent\textbf{Training setup.}
We fine-tune a Gemma-2-2B\footnote{\url{https://huggingface.co/google/gemma-2-2b}} encoder with LoRA on a single NVIDIA V100 (32GB) using mixed precision.
On TREC DL, we construct 20{,}000 $(q,d^{+},d^{-})$ triplets; negatives are uniformly sampled from non-positives.
On MLDR-zh, we follow the official training files and sample one positive and one negative per query (10{,}000 triplets total), where the negative is chosen as the first labeled negative.
We optimize the pairwise hinge loss with margin $m{=}10$ (following RepLLaMA, we rescale similarity scores by a factor of $0.01$; the margin is set in the same scaled score space).
Unless stated otherwise, we set LoRA rank $r{=}32$ and $\alpha{=}64$, use learning rate $5{\times}10^{-5}$, batch size $1$ with gradient accumulation of $4$, and cap the query length at 32 tokens.

\noindent\textbf{Segmentation and block budget.}
Documents are segmented into 63-token blocks.
Unless noted, we train with top-$k{=}20$ blocks per document (approximately 1{,}260 tokens) and evaluate by varying $k \in \{20,30,50,65\}$ under a 4k-token budget (and additionally $k{=}130$ on MLDR-zh when reporting extended-budget results).

\noindent\textbf{Indexing and reranking.}
Inference proceeds in two phases.
\emph{Offline}, we encode each document into block representations and store up to the first $n$ blocks per document, where $n$ is determined by the evaluation budget.
\emph{Online}, given a query representation, we score all stored blocks of a candidate document, select the top-$k$ blocks, and aggregate their scores into a document score.
We use monotonically decreasing weights over the top-$k$ block scores for aggregation.

\noindent\textbf{Reranking protocol.}
We study \emph{long-document reranking} under fixed candidate sets:
for TREC DL, we rerank the {top-100} candidates per query from the official lists;
for MLDR-zh, we rerank the provided candidate sets (one positive plus seven negatives per query).
All methods are evaluated on the same candidates, and we match the document token budget whenever applicable.

\noindent\textbf{Baseline input budgets.}
For RepLLaMA, we use the first 2{,}048 document tokens for training/evaluation due to GPU constraints, and additionally report 4{,}096-token evaluation where applicable.
For BERT-dot and ColBERT, we use \texttt{bert-base-uncased} (English) and \texttt{bert-base-chinese} (MLDR-zh) with the first 512 document tokens; we train them with learning rate $1{\times}10^{-5}$, batch size 8, and mixed precision, without LoRA.
For long-context embedding models, we evaluate 
E5-Base-4K with up to 4{,}096 tokens using official checkpoints; BGE-M3 is evaluated with 4{,}096 and 8{,}192 tokens.
For segmentation-based rerankers, KeyB(BERT)$_{\text{BM25}}$ operates on full documents via block segmentation and aggregation (i.e., without a global document-level truncation), while IDCM is evaluated with a 4{,}096-token maximum input length.

\noindent\textbf{Top-$k$ Interaction Refinement (TIR).}
TIR refines the aggregation weights of the top-ranked blocks via query-conditioned attention and applies a residual score correction.
We set a lightweight head with $d{=}256$, $\tau_{\mathrm{att}}{=}0.07$, and $\gamma{=}0.3$, selected on the development set and then fixed for all experiments.

\noindent\textbf{Efficiency and footprint.}
We report \emph{online} end-to-end reranking latency (ms/query) including tokenization, host-to-device transfer, inference, and output writing, while excluding model loading and offline embedding precomputation.
We also estimate the embedding footprint analytically (bytes/doc) assuming fp16 storage.

\subsection{Main Results and Analysis}
\label{sec:main-results}

Tables~\ref{tab:dl19}--\ref{tab:mldrzh} summarize the main results on DL'19, DL'22, DL'23, and MLDR-zh under a fixed-candidate reranking protocol.
For compactness, we report our primary 4k-budget variants (65 blocks) on TREC DL; additional budget sweeps are analyzed in the ablation section (Sec.~\ref{sec:ablation-k}).
Unless otherwise stated, superscripts $^{b}$ and $^{t}$ indicate statistically significant differences from \textbf{BReps-65} and \textbf{BReps-TIR-65}, respectively (paired $t$-test on per-query scores, $p<0.05$).

\begin{table}[htbp]
\centering
\small
\caption{TREC 2019 DL (Document Ranking).
Superscripts $^{b}$ and $^{t}$ denote statistically significant difference
from \textbf{BReps-65} and \textbf{BReps-TIR-65}, respectively
(paired $t$-test on per-query scores, $p<0.05$).}
\label{tab:dl19}
\begin{tabular}{l l l l}
\toprule
\multicolumn{1}{c}{Model} & \multicolumn{1}{c}{Context} & \multicolumn{1}{c}{NDCG@10} & \multicolumn{1}{c}{MAP} \\
\midrule
\multicolumn{4}{l}{\textit{\textbf{Baseline models}}} \\
\addlinespace[2pt]
BM25                 & Full document           & 0.488 & 0.234 \\
BERT-dot             & 110M, 512 tokens        & 0.638 & 0.241 \\
ColBERT              & 110M, 512 tokens        & 0.642 & 0.268 \\
\midrule
\multicolumn{4}{l}{\textit{\textbf{PLMs and LLMs for long documents}}} \\
\addlinespace[2pt]
E5-Base-4K           & 110M, 4096 tokens       & 0.643$^{t}$ & 0.274 \\
BGE-M3               & 560M, 4096 tokens       & 0.639$^{t}$ & 0.261$^{bt}$ \\
BGE-M3               & 560M, 8192 tokens       & 0.632$^{bt}$ & 0.258$^{bt}$ \\
RepLLaMA             & 2B, 2k tokens           & 0.658$^{t}$ & 0.269 \\
RepLLaMA             & 2B, 4k tokens           & 0.657$^{t}$ & 0.269 \\
IDCM & 4096 tokens & 0.681$^{t}$&0.277 \\
KeyB(BERT)$_{BM25}$ & full doc & 0.683 & \textbf{0.281} \\
\midrule
\multicolumn{4}{l}{\textit{\textbf{Proposed: BReps and BReps-TIR}}} \\
BReps-65 blocks      & $\leq$4095 tokens       & 0.683 & 0.274 \\
BReps-TIR-65 blocks      & $\leq$4095 tokens       & \textbf{0.700} & 0.275 \\
\bottomrule
\end{tabular}
\end{table}

\subsubsection{TREC 2019 DL}
Table~\ref{tab:dl19} reports results on DL'19 (NDCG@10, MAP).

\textbf{Baselines.}
BM25 is the weakest baseline, while BERT-dot and ColBERT substantially improve effectiveness, confirming the benefit of neural matching for long documents.

\textbf{Long-context embedding baselines.}
E5-Base-4K and BGE-M3 are competitive but remain below the strongest rerankers.
Notably, extending BGE-M3 from 4k to 8k tokens degrades both NDCG@10 and MAP, suggesting that simply increasing context can introduce noise.
RepLLaMA benefits from long-context encoding (2k/4k tokens) and achieves strong NDCG@10, but it is still significantly below our best model ($^{t}$).

\textbf{Segmentation-and-aggregation baselines.}
IDCM and KeyB$_{BM25}$ are strong long-document rerankers in this setting: IDCM attains 0.681 NDCG@10, and KeyB achieves the best MAP (0.281).
However, \textbf{BReps-TIR-65} achieves the best overall NDCG@10 (\textbf{0.700}) with competitive MAP (0.275), and significantly outperforms most baselines in NDCG@10 ($p<0.05$ as marked).

\begin{table}[htbp]
\centering
\small
\caption{TREC 2022 DL (Document Ranking).
Superscripts $^{b}$ and $^{t}$ denote statistically significant difference
from \textbf{BReps-65} and \textbf{BReps-TIR-65}, respectively
(paired $t$-test on per-query scores, $p<0.05$).}
\label{tab:dl22}
\begin{tabular}{l l l l}
\toprule
\multicolumn{1}{c}{Model} & \multicolumn{1}{c}{Context} & \multicolumn{1}{c}{NDCG@10} & \multicolumn{1}{c}{MAP} \\
\midrule
\multicolumn{4}{l}{\textit{\textbf{Baseline models}}} \\
\addlinespace[2pt]
BM25                 & Full document           & 0.299 & 0.080 \\
BERT-dot             & 110M, 512 tokens        & 0.324 & 0.088 \\
ColBERT              & 110M, 512 tokens        & 0.349 & 0.096 \\
\midrule
\multicolumn{4}{l}{\textit{\textbf{PLMs and LLMs for long documents}}} \\
\addlinespace[2pt]
E5-Base-4K           & 110M, 4096 tokens       & 0.399$^{t}$ & 0.107$^{t}$ \\
BGE-M3               & 560M, 4096 tokens       & 0.401$^{t}$ & 0.107$^{t}$ \\
BGE-M3               & 560M, 8192 tokens       & 0.393$^{t}$ & 0.106$^{t}$ \\
RepLLaMA             & 2B, 2k tokens           & 0.374$^{t}$ & 0.099$^{t}$ \\
RepLLaMA             & 2B, 4k tokens           & 0.376$^{t}$ & 0.099$^{t}$ \\
IDCM & 4096 tokens & 0.402$^{t}$&0.103$^{t}$ \\
KeyB(BERT)$_{BM25}$ & full doc & 0.362$^{bt}$ & 0.099$^{t}$ \\
\midrule
\multicolumn{4}{l}{\textit{\textbf{Proposed: BReps and BReps-TIR}}} \\
BReps-65      & $\leq$4095 tokens       & 0.398$^{t}$ & 0.106$^{t}$ \\
BReps-TIR-65      & $\leq$4095 tokens       & \textbf{0.436} & \textbf{0.115} \\
\bottomrule
\end{tabular}
\end{table}

\begin{table}[htbp]
\centering
\small
\caption{TREC 2023 DL (Document Ranking).
Superscripts $^{b}$ and $^{t}$ denote statistically significant difference
from \textbf{BReps-65} and \textbf{BReps-TIR-65}, respectively
(paired $t$-test on per-query scores, $p<0.05$).}
\label{tab:dl23}
\begin{tabular}{l l l l}
\toprule
\multicolumn{1}{c}{Model} & \multicolumn{1}{c}{Context} & \multicolumn{1}{c}{NDCG@10} & \multicolumn{1}{c}{MAP} \\
\midrule
\multicolumn{4}{l}{\textit{\textbf{Baseline models}}} \\
\addlinespace[2pt]
BM25                 & Full document           & 0.295 & 0.105 \\
BERT-dot             & 110M, 512 tokens        & 0.296 & 0.101 \\
ColBERT              & 110M, 512 tokens        & 0.315 & 0.111 \\
\midrule
\multicolumn{4}{l}{\textit{\textbf{PLMs and LLMs for long documents}}} \\
\addlinespace[2pt]
E5-Base-4K           & 110M, 4096 tokens       & 0.391$^{t}$ & 0.133$^{t}$ \\
BGE-M3               & 560M, 4096 tokens       & 0.415$^{t}$ & 0.141 \\
BGE-M3               & 560M, 8192 tokens       & 0.400$^{t}$ & 0.136$^{t}$ \\
RepLLaMA             & 2B, 2k tokens           & 0.346$^{bt}$ & 0.120$^{bt}$ \\
RepLLaMA             & 2B, 4k tokens           & 0.341$^{bt}$ & 0.120$^{bt}$ \\
IDCM & 4096 tokens & 0.400$^{t}$&0.132$^{t}$ \\
KeyB(BERT)$_{BM25}$ & full doc  & 0.352$^{bt}$ & 0.123$^{bt}$ \\
\midrule
\multicolumn{4}{l}{\textit{\textbf{Proposed: BReps and BReps-TIR}}} \\
\addlinespace[2pt]
BReps-65      & $\leq$4095 tokens       & 0.395$^{t}$ & 0.133$^{t}$ \\
BReps-TIR-65      & $\leq$4095 tokens       & \textbf{0.451} & \textbf{0.147} \\
\bottomrule
\end{tabular}
\end{table}

\subsubsection{TREC 2022 DL and TREC 2023 DL}
Tables~\ref{tab:dl22} and \ref{tab:dl23} report results on MS~MARCO v2 based DL'22 and DL'23 (NDCG@10, MAP).

\textbf{Baselines.}
BM25 remains weak; BERT-dot/ColBERT provide consistent gains but are clearly below long-context embedding models and our methods.

\textbf{Long-context embedding baselines.}
E5-Base-4K and BGE-M3 (4k) are strong on both datasets, while BGE-M3 at 8k does not consistently improve and can drop, echoing DL'19.
RepLLaMA underperforms these embedding baselines on MS~MARCO v2 and is significantly below our best results.

\textbf{Segmentation-and-aggregation baselines.}
IDCM is competitive among non-LLM long-document rerankers (e.g., DL'22: 0.402 NDCG@10; DL'23: 0.400), yet it is significantly below \textbf{BReps-TIR-65} on both datasets ($^{t}$).
KeyB(BERT)$_{BM25}$ performs notably worse on v2 (DL'22/DL'23) and is also significantly below our models.

\textbf{Proposed methods.}
With the same 4k budget, \textbf{BReps-TIR-65} achieves the best effectiveness on both DL'22 (\textbf{0.436}/\textbf{0.115}) and DL'23 (\textbf{0.451}/\textbf{0.147}), surpassing all baselines.
Compared to BReps-65, TIR provides a clear gain at the same input budget, supporting the value of lightweight setwise refinement over the top blocks.

\begin{table}[htbp]
\centering
\small
\caption{Results on MLDR-zh (Document Ranking).
For models truncated at 65 blocks, superscripts $^{b}$ and $^{t}$ denote statistically significant difference
from \textbf{BReps-65} and \textbf{BReps-TIR-65}, respectively.
For comparisons at 8192 tokens (and KeyB) and 130 blocks, significance is measured against \textbf{BReps-130} and \textbf{BReps-TIR-130} (paired $t$-test on per-query scores, $p<0.05$).}
\label{tab:mldrzh}
\resizebox{\linewidth}{!}{
\begin{tabular}{l l l l l}
\toprule
\multicolumn{1}{c}{Model} & \multicolumn{1}{c}{Context} & \multicolumn{1}{c}{P@1} & \multicolumn{1}{c}{MAP} & \multicolumn{1}{c}{NDCG@8} \\
\midrule
\multicolumn{5}{l}{\textit{\textbf{Baseline models}}} \\
\addlinespace[2pt]
BM25                 & Full document           & 0.201 & 0.259 & 0.277 \\
BERT-dot             & 110M, 512 tokens        & 0.600 & 0.733 & 0.798 \\
ColBERT              & 110M, 512 tokens        & 0.586 & 0.723 & 0.790 \\
\midrule
\multicolumn{5}{l}{\textit{\textbf{PLMs and LLMs for long documents}}} \\
\addlinespace[2pt]
E5-Base-4K           & 110M, 4096 tokens       & 0.406$^{bt}$ & 0.576$^{bt}$ & 0.607$^{bt}$ \\
BGE-M3               & 560M, 4096 tokens       & 0.878$^{b}$  & 0.926$^{b}$  & 0.945$^{b}$ \\
BGE-M3               & 560M, 8192 tokens       & 0.906$^{t}$  & 0.945 & 0.959 \\
RepLLaMA             & 2B, 2k tokens           & 0.666$^{bt}$ & 0.775$^{bt}$ & 0.830$^{bt}$ \\
RepLLaMA             & 2B, 4k tokens           & 0.718$^{bt}$ & 0.815$^{bt}$ & 0.860$^{bt}$ \\
KeyB(BERT)$_{BM25}$ & full doc & 0.423$^{bt}$ & 0.586$^{bt}$ & 0.685$^{bt}$ \\
\midrule
\multicolumn{5}{l}{\textit{\textbf{Proposed: BReps and BReps-TIR}}} \\
\addlinespace[2pt]
BReps-65             & 2B, $\leq$4095 tokens   & 0.849 & 0.901$^{t}$ & 0.925$^{t}$ \\
BReps-TIR-65         & 2B, $\leq$4095 tokens   & 0.869 & 0.918 & 0.939 \\
BReps-130            & 2B, $\leq$8190 tokens   & 0.920 & 0.945 & 0.959 \\
BReps-TIR-130        & 2B, $\leq$8190 tokens   & \textbf{0.928} & \textbf{0.955} & \textbf{0.966} \\
\bottomrule
\end{tabular}
}
\end{table}

\subsubsection{MLDR-zh}
Table~\ref{tab:mldrzh} reports results on MLDR-zh (P@1, MAP, NDCG@8), where each test query is ranked over the official eight candidates (one positive and seven negatives).

\textbf{Baselines.}
BM25 is weak, while BERT-dot/ColBERT yield strong improvements.
Among long-context embeddings, BGE-M3 is the strongest baseline and benefits from larger context, whereas RepLLaMA and KeyB(BERT)$_{BM25}$ lag behind substantially.

\textbf{Proposed methods.}
At the 4k budget (65 blocks), BReps already achieves strong effectiveness (0.849/0.901/0.925), and TIR further improves to 0.869/0.918/0.939.
With a larger 8k budget (130 blocks), \textbf{BReps-TIR-130} attains the best overall results (\textbf{0.928}/\textbf{0.955}/\textbf{0.966}), matching or exceeding BGE-M3 at 8k.
(We do not include IDCM on MLDR-zh due to the lack of an official Chinese checkpoint.)

\subsection{Visualization of Representations}
\begin{figure}[htbp]
  \centering
  \includegraphics[width=\columnwidth]{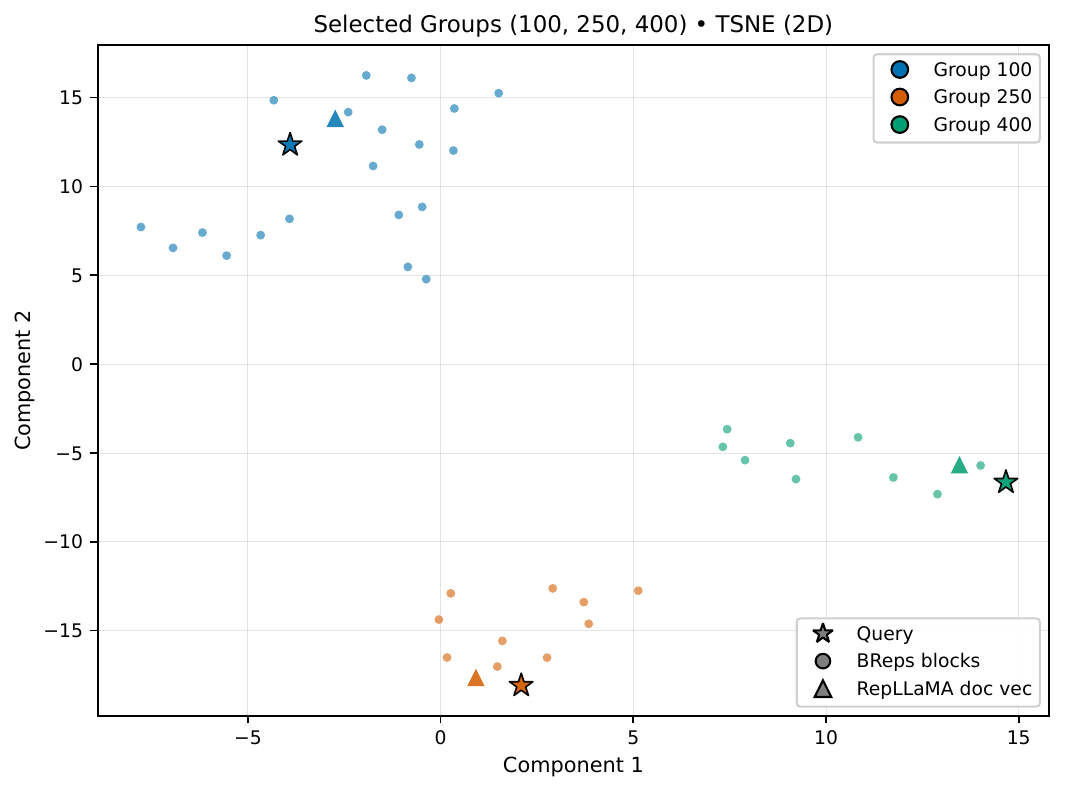}
  \caption{Case study of three query–document pairs (sampled in MS MARCO v1 doc). 
  Representations are projected from the original embedding space into 2D with t-SNE. 
  Multiple BReps block embeddings lie closer and broader to the query than RepLLaMA’s single document embedding, and they spread across clusters, showing finer-grained and complementary matching signals. }
  \label{fig:case-study}
\end{figure}

Beyond numerical improvements, Fig.~\ref{fig:case-study} illustrates an example of how BReps differs from RepLLaMA in the representation space. We first randomly compute 500 query-relevant document pairs' representations offline from MS MARCO v1 training set. Then show three groups' representations. 
RepLLaMA produces a single coarse document embedding, BReps yields multiple block-level embeddings. 
When projected to 2D with t-SNE \cite{van2008visualizing}, several BReps embeddings appear closer to the query and distributed broader across clusters, indicating finer-grained and complementary relevance signals.

\subsection{Efficiency Analysis}
\label{sec:efficiency}

Figure~\ref{fig:ndcg-length} summarizes the overall effectiveness--efficiency trade-off on TREC DL'19 under the \emph{online} reranking setting.
We further provide a breakdown of efficiency from two complementary aspects: (i) offline representation computation, and (ii) online end-to-end reranking latency.

\subsubsection{Offline Representation Computation}
\label{sec:eff-offline}

\begin{figure}[htbp]
\centering
\includegraphics[width=\columnwidth]{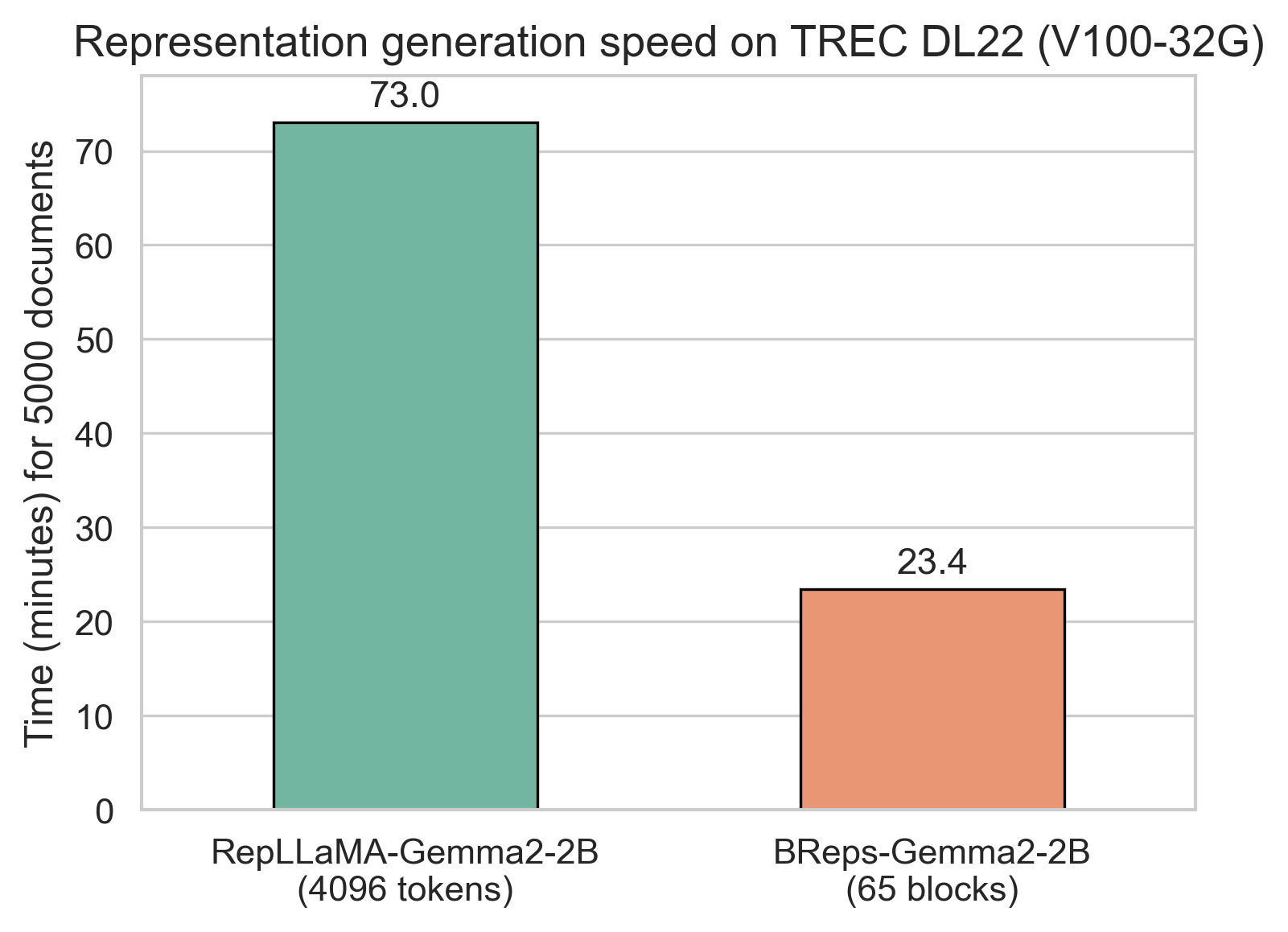}
\caption{Time to compute representations for 5,000 TREC DL'22 documents on one V100-32GB GPU. RepLLaMA encodes up to 4,096 tokens; BReps encodes up to 65 blocks.}
\label{fig-latency}
\end{figure}

In addition to effectiveness, we measure the time required to compute and materialize document representations for 5000 documents from TREC DL'22 (chosen because MS MARCO v2 documents are substantially longer than v1).
We run all experiments on a single V100-32GB GPU.
For RepLLaMA, we encode up to the first 4096 tokens per document; for BReps, we encode up to the first {65} blocks (under the same 4k-token budget), as illustrated in Fig.~\ref{fig-latency}.

Using Gemma2-2B, RepLLaMA requires \textbf{73.0} minutes to compute representations, whereas BReps requires only \textbf{23.4} minutes under the same budget, reducing the computation time by \textbf{67.9\%}.
This improvement is expected because self-attention cost grows superlinearly with sequence length~\cite{vaswani2017attention}; by processing a long document as multiple shorter blocks, BReps avoids repeatedly attending over long sequences and thus reduces offline computation overhead.

\subsubsection{Online End-to-End Reranking Speed}
\label{sec:eff-online}

We also measure \emph{online} end-to-end reranking latency (ms/query) on TREC DL'19 when reranking the top-100 candidates per query.
Following Fig.~\ref{fig:ndcg-length}, latency includes tokenization, host-to-device transfer, inference, and output writing, but excludes model loading and any offline precomputation.
On a single NVIDIA A100 GPU, BReps takes \textbf{239.63} ms/query to rerank 100 documents, while adding TIR increases latency to \textbf{289.91} ms/query (a $1.21\times$ overhead).
In contrast, segmentation-and-cross-encoder baselines are substantially slower under the same protocol: KeyB(BERT)$_{\text{BM25}}$ takes \textbf{1,189.81} ms/query for 100 documents (about $5.0\times$ slower than BReps), and IDCM with a 4,096-token input cap takes \textbf{1,291.75} ms/query (about $5.4\times$ slower).
Overall, BReps achieves favorable efficiency both in offline representation computation (Fig.~\ref{fig-latency}) and in online reranking (Fig.~\ref{fig:ndcg-length}), while TIR provides effectiveness gains with only a modest additional cost.

\subsubsection{Embedding Footprint (Storage Cost)}
\label{sec:footprint}

Beyond latency, we estimate the \emph{storage footprint} of document representations for deployment.
Assuming fp16 storage, each scalar takes 2 bytes.
For a single-vector reranker with embedding dimension $d$, the per-document storage is
\begin{equation}
\mathrm{Bytes/doc} = 2d. \nonumber
\end{equation}
For BReps, each document is represented by up to $k$ block embeddings of the same dimension $d$, yielding
\begin{equation}
\mathrm{Bytes/doc} = 2kd. \nonumber
\end{equation}
For late-interaction (ColBERT-style) rerankers that store token-level embeddings, the footprint scales with the number of stored tokens:
storing $L$ token embeddings per document costs approximately $\mathrm{Bytes/doc} \approx 2Ld$ (fp16).

For example, with $d{=}256$ and $k{=}65$ (4k budget), BReps stores $2{\times}65{\times}256 \approx 33.3$KB per document, while a single-vector representation stores $2{\times}256 = 512$B.
Compared with token-level late-interaction representations (which scale linearly with the stored token budget $L$), block-level storage provides a more controllable footprint by operating on a small number of blocks ($k$) per document.

This analytical estimate isolates the representation storage cost. In contrast, cross-encoder-style long-document rerankers such as IDCM and KeyB primarily incur cost at query time by scoring query--text pairs, rather than storing large document-side embedding arrays for fast similarity scoring.

\subsection{Ablation Studies and Hyperparameter Sensitivity}
\label{sec:furtherInvest}

We further investigate how key design choices affect the behavior of BReps, including segmentation, aggregation, the number of selected blocks, and the Top-$k$ Interaction Refinement (TIR) module.

\subsubsection{Segmentation Strategy}
\label{sec:ablation-seg}

We adopt the dynamic-programming segmentation method of CogLTX~\cite{ding2020cogltx}, which aims to form semantically coherent blocks while respecting a token budget.
To assess whether a simpler strategy is sufficient, we replace CogLTX with \emph{fixed-length} segmentation and vary the block length on TREC DL'19, keeping all other settings identical to the default BReps configuration in Table~\ref{tab:dl19}.
Table~\ref{tab:dl19DiffLength} reports the results.

\begin{table}[htbp]
  \centering
  \caption{Results of different fixed block lengths with BReps compared with the default (CogLTX) segmentation on TREC DL'19.}
  \scalebox{0.95}{

\begin{tabular}{@{}llllllll@{}}
\toprule
block length &NDCG@10 &MAP  \\ \midrule
default (CogLTX) & 0.684 & 0.272 \\
50 &0.656 & 0.274 \\
60 &0.672 & 0.270 \\
66 &0.675 & 0.271 \\
80 &0.665 & 0.270 \\
100 &0.676 & 0.267 \\
 \bottomrule
\end{tabular}
  }
  \label{tab:dl19DiffLength}
\end{table}

Overall, fixed-length segmentation is consistently worse than CogLTX on NDCG@10, and typically also worse on MAP.
While a block length of 50 slightly improves MAP, it yields a noticeable drop in NDCG@10, indicating that naive fixed-length boundaries can fragment relevant evidence and dilute high-precision signals.
These results highlight the importance of semantically informed segmentation for block-based long-document reranking.

\subsubsection{Aggregation Function and Weights}
\label{sec:ablation-agg}

BReps aggregates the top-$k$ block scores using \emph{descending} weights that emphasize the most salient blocks, following prior work~\cite{li2022bert}.
We test two alternatives on TREC DL'19: (i) \emph{average} weights, and (ii) \emph{learned} weights.
For a controlled comparison, we train the same BReps and BReps-TIR models as in Table~\ref{tab:dl19} using the first 20 blocks per document.

\begin{table}[htbp]
  \centering
  \caption{Aggregation weight ablations on TREC DL'19. We compare descending weights (default) with average weights and learned weights.}
  \scalebox{0.9}{
    
\begin{tabular}{@{}llll@{}}
\toprule
& NDCG@10 &MAP  \\ \midrule
BReps-20Blocks-default & 0.684 &0.272 \\
BReps-20Blocks-avgWeight & 0.671&0.271 \\
BReps-20Blocks-LearnWeight & 0.664&0.269 \\
BReps-TIR-20Blocks-default & 0.691 &0.273 \\
BReps-TIR-20Blocks-avgWeight & 0.682&0.271 \\
BReps-TIR-20Blocks-LearnWeight & 0.679&0.274 \\
 \bottomrule
\end{tabular}

  }
  \label{tab:dl19AvgWeight}
\end{table}

Table~\ref{tab:dl19AvgWeight} shows that average weights consistently reduce both NDCG@10 and MAP, confirming that uniformly averaging across blocks is suboptimal for long documents.
Learning weights does not improve over the default scheme either: for BReps, both metrics drop; for BReps-TIR, MAP slightly improves but at the cost of a noticeable decrease in NDCG@10.
Overall, descending weights provide the best effectiveness--robustness trade-off and are used by default.

\paragraph{Pooling baselines vs.\ weighted Top-$k$ aggregation.}
To isolate the effect of the aggregation function, we further replace our weighted Top-$k$ aggregation with two standard pooling baselines over the \emph{same} Gemma2-2B block scores: AvgP (mean pooling over blocks) and MaxP (max pooling).
All methods use the same block encoder and the same block pool (up to 65 blocks) under the 4k-token budget; only the aggregation differs.
Table~\ref{tab:ablation_pooling} reports the results on TREC DL'19.

\begin{table}[htbp]
\centering
\caption{Ablation on aggregation functions on TREC DL'19.}
\begin{tabular}{lcc}
\toprule
Aggregation (Gemma2-2B block scores) & nDCG@10 & MAP \\
\midrule
AvgP (mean pooling) & 0.553 & 0.249 \\
MaxP (max pooling)  & 0.669 & 0.269 \\
Weighted Top-$k$ (BReps) & \textbf{0.683} & \textbf{0.274} \\
\bottomrule
\end{tabular}
\label{tab:ablation_pooling}
\end{table}

AvgP performs poorly, indicating that averaging over many blocks dilutes relevance signals with substantial noise.
MaxP is considerably stronger but still falls short of weighted Top-$k$, suggesting that relying on a single best-matching block is less robust than softly weighting multiple high-saliency blocks.

\subsubsection{Number of Selected Blocks $k$}
\label{sec:ablation-topk}

\begin{figure}[htbp]
  \centering
  \includegraphics[width=0.95\linewidth]{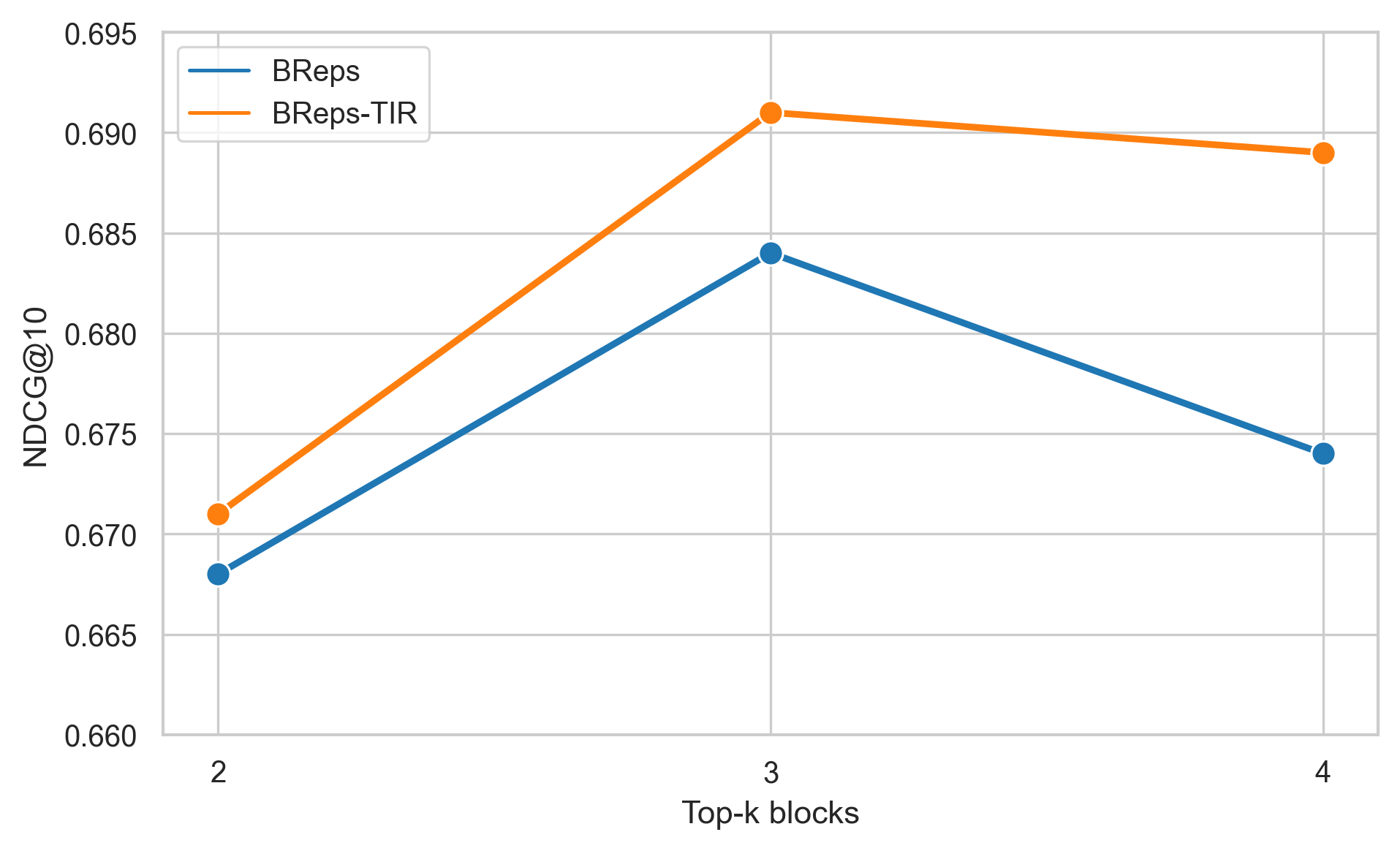}
  \caption{Effect of the number of selected blocks $k$ on NDCG@10 (TREC DL'19).}
  \label{fig:ablation_topk}
\end{figure}

We next study how many top-scoring blocks should be aggregated.
Keeping the same block pool, scoring function, and descending-weight aggregation, we vary the number of selected blocks $k$.
Figure~\ref{fig:ablation_topk} reports NDCG@10 on TREC DL'19.
Selecting too few blocks under-utilizes relevant evidence, whereas overly large $k$ introduces noise from marginal blocks; the default setting achieves the best trade-off.

\subsubsection{Effect of Block Budget and TIR}
\label{sec:ablation-k}

\begin{figure}[t]
  \centering
  \includegraphics[width=\columnwidth]{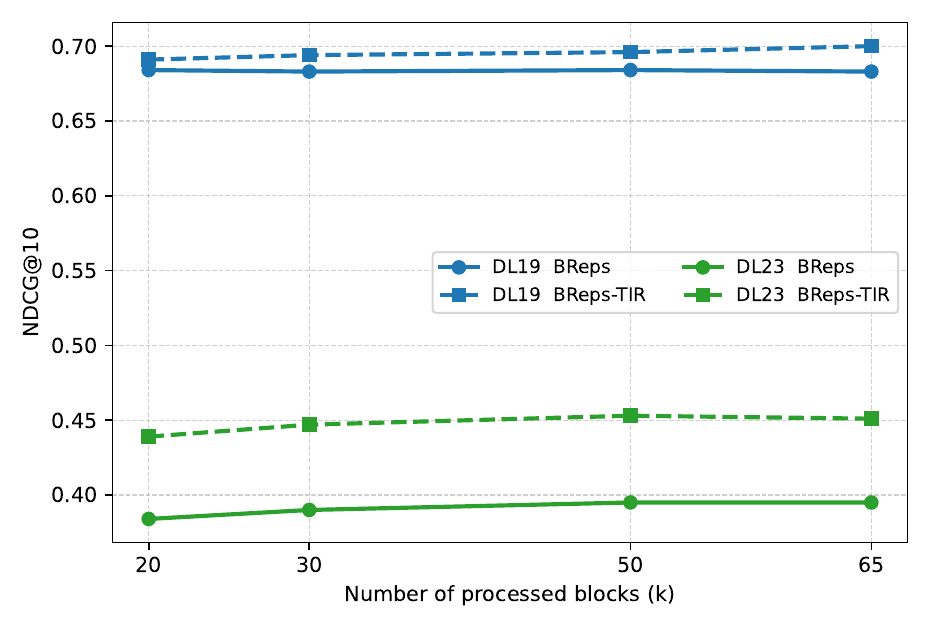}
  \caption{Effect of block budget $k$ on NDCG@10 (DL'19 and DL'23). Solid: BReps; dashed: BReps-TIR.}
  \label{fig:ablation-k}
\end{figure}

To keep the main tables concise, we report only the 4k-budget setting ($k{=}65$) in the main results.
Here we vary the number of processed blocks $k$ and evaluate NDCG@10 on DL'19 and DL'23.
Figure~\ref{fig:ablation-k} shows that increasing $k$ generally improves effectiveness, and TIR consistently outperforms BReps at matched budgets.

\subsubsection{Ablation of Top-$k$ Interaction Refinement (TIR)}
\label{sec:ablation-tir}

We isolate the contribution of each TIR component by removing query$\rightarrow$block attention (i.e., no setwise context), removing the score-gating MLP, and examining the resulting effectiveness changes.
Table~\ref{tab:tir-ablation-main} reports the core comparisons on TREC DL'19: the full TIR is consistently best; removing attention or score-gating yields clear drops in NDCG@10 (and similar trends for MAP, omitted due to space).

\begin{table}[ht]
\centering
\caption{Core ablation on TREC DL'19 (NDCG@10) for the TIR module. $\Delta$ is relative to the default.}
\label{tab:tir-ablation-main}
\small
\begin{tabular}{lcc}
\toprule
\textbf{Variant} & \textbf{NDCG@10} & $\boldsymbol{\Delta}$ \\
\midrule
Default (full TIR)   & 0.691 & -- \\
w/o Attention        & 0.676 & $-0.015$ \\
w/o Score-Gate       & 0.672 & $-0.019$ \\
\bottomrule
\end{tabular}
\end{table}

\subsubsection{TIR Parameter Size}
\label{sec:tir-size}

Finally, we measure the parameter size of TIR using PyTorch.
TIR contains \textbf{3.03M} parameters in total, which is nearly two orders of magnitude fewer than BERT-base (110M) and several thousand times smaller than typical LLMs (e.g., LLaMA-7B with 7{,}000M parameters).
This confirms that TIR is a lightweight component that can be integrated into practical long-document reranking pipelines with minimal overhead.

\section{Conclusion}
\label{sec:conclusion}

We presented \textbf{BReps}, an \emph{efficient long-document reranking} framework under a fixed-candidate setting.
Instead of compressing a long document into a single coarse vector, BReps represents each document with multiple \emph{block-level} embeddings and aggregates the top-$k$ query--block similarities to produce a document score.
To further calibrate the selected blocks, we introduced \textbf{Top-$k$ Interaction Refinement (TIR)}, a lightweight query-conditioned setwise module that models dependencies among top blocks and applies bounded residual corrections with modest query-time overhead.

Across long-document reranking benchmarks (TREC DL'19/22/23 and MLDR-zh), \textbf{BReps} and \textbf{BReps-TIR} consistently improve over single-vector LLM encoders such as RepLLaMA, and achieve a favorable effectiveness--efficiency trade-off compared with segmentation-and-cross-encoder baselines (e.g., IDCM and KeyB) under the same fixed-candidate protocol.
These results suggest that preserving fine-grained block evidence, together with lightweight setwise refinement over only the most salient blocks, is an effective and practical alternative to either truncation-based single-vector encoding or expensive interaction-heavy rerankers.

Future work includes extending block representations to earlier-stage retrieval and ANN indexing at scale, exploring adaptive block budgeting/selection conditioned on query difficulty, and improving multilingual coverage and robustness for extremely long documents.


\bibliographystyle{ACM-Reference-Format}
\bibliography{myRef}

\end{document}